\documentclass[twocolumn]{aa}
\usepackage{graphicx}
\usepackage{txfonts}
\begin{document}
   \title{The UV luminosity function of nearby clusters of galaxies}

   \author{L.Cortese\inst{1}, G.Gavazzi\inst{1}, A.Boselli\inst{2}, J.Iglesias-Paramo\inst{2}, J. Donas\inst{2} \and B. Milliard\inst{2}}

   \offprints{L.Cortese}

   \institute{Universit\'{a} degli Studi di Milano-Bicocca, P.zza della Scienza 3, 20126 Milano, Italy.\\
              \email{Luca.Cortese@mib.infn.it; Giuseppe.Gavazzi@mib.infn.it}
         \and
             Laboratoire d'Astrophysique de Marseille, BP8, Traverse du Siphon - F-13376 Marseille Cedex 12, France.\\
             \email{alessandro.boselli@oamp.fr; jorge.iglesias@oamp.fr}
             }

   \date{Received 17 July 2003 | Accepted 23 September 2003}

   \abstract{We present the UV composite luminosity function for galaxies in the Virgo, Coma and Abell 1367 clusters.
   	     The luminosity function (LF) is well fitted by a Schechter function with 
	     $M_{\rm UV}^* - 5~ \log~h_{\rm 75}$= $-$20.75 $\pm$ 0.40 and $\alpha$ = $-$1.50 $\pm$ 0.10 
	     and does not differ significantly from the local UV luminosity function of the field.
	     This result is in agreement with recent studies carried out in the $\rm H\alpha$ and B-bands
	     which find no difference between the LFs of star forming galaxies in clusters and in the field.
	     This indicates that, whatever mechanisms are responsible for quenching the star formation in clusters, they 
	     influence similarly the giant and the dwarf populations, 
	     leaving the shape of the LF unchanged and only modifying its normalization.

   }

\titlerunning{The UV luminosity function of nearby clusters}
\authorrunning{L.Cortese et al.}

   \maketitle
%

\section{Introduction}
The study of the galaxy luminosity function (hereafter LF) provides us with 
a fundamental tool for testing theories of galaxy formation and for
reconstructing their evolution to the present. 
Accurate measurements of the LF in the local field, in  nearby clusters and in clusters at progressively 
high redshift can improve our knowledge of galaxy evolution and on the role played by the environment in regulating the star formation 
activity of galaxies.
Recent studies, based on H$\alpha$ (Iglesias-Paramo et al. \cite{jorge02}) and B band observations (De Propris et al. \cite{depropris}), 
find no significant differences between the LF of star forming galaxies in the field and in clusters.\\
An excellent tool to identify and quantify the star formation activity is represented by the ultraviolet emission. 
Although the shape of local field UV LF (Sullivan et al. \cite{sullivan}) is well determined, there is still a fair amount of uncertainty on the UV luminosity 
function of clusters. Its slope is undetermined due to the insufficient knowledge of the background counts (Cortese et al. \cite{cortese}). 
Andreon (\cite{andreon}) proposed a very steep faint end ($\alpha \sim -2.0,-2.2$), significantly different from the field LF ($\alpha \sim -1.5$).
However Cortese et al.(\cite{cortese}) pointed out that this steep slope is likely caused by an underestimation of the density of background galaxies and 
proposed a flatter faint-end slope ($\alpha \sim -1.35 \pm 0.20$). 
Unfortunately the statistical uncertainty was too high for making reliable comparisons between the cluster and the field LFs.
In this paper we re-compute the cluster UV luminosity function with two major improvements over previous determinations.
We increase the redshift completeness of the UV selected sample using new spectroscopic observations of Coma and Abell 1367 (Cortese et al., in preparation), 
and compute for the first time the UV LF of the Virgo cluster.
These improvements are not sufficient to constrain the LF of each individual cluster, however the UV composite luminosity function, 
constructed for the first time in this paper can be significantly compared with that of the field.
Doing so we try anticipating one of the main goals of the Galaxy Evolution Explorer (GALEX) which, within 
one year, will shed light on the UV properties of galaxies and their environmental dependences.\\
We assume a distance modulus $\mu$= 31.15 for the Virgo cluster (Gavazzi et al. \cite{gav99a}), 
$\mu$=34.80 for Abell 1367 and $\mu$=34.91 for the Coma cluster (Gavazzi et al. \cite{gav99b}), 
corresponding to a Hubble constant $H_0 = 75 \rm km~s^{-1}~Mpc^{-1}$.

\section{The Data}
The sample analyzed in this work comprises the UV sources detected in Virgo, Coma and 
Abell 1367 clusters by the FOCA (Milliard et al. \cite{milliard}) and FAUST (Lampton et al. \cite{lampton}) experiments.
The FOCA balloon-borne wide field UV camera ($\rm \lambda = 2000\AA$; $\rm \Delta\lambda=150\AA$) observed  
$\sim$ 3~ square degrees ($\sim 8~ \rm Mpc^{2}$) in the Abell 1367 (unpublished data) and Coma clusters (Donas et al. \cite{donas95}) 
and $\sim$ 12~ square degrees ($\sim 1~ \rm Mpc^{2}$) in the Virgo cluster 
(data are taken from the extragalactic database GOLDMine, Gavazzi et al. \cite{gavazzi03}).
The FOCA observations of Virgo are not sufficient to compile a complete catalog: no sources brighter than $m_{\rm UV}\sim12.2$ were detected
due to the small area covered. We thus complement the UV database with the wide field observations performed by the FAUST 
space experiment ($\rm \lambda = 1650\AA$; $\rm \Delta\lambda=250\AA$) in the Virgo direction (Deharveng et al. \cite{deharveng94}), 
covering $\sim$ 100~ square degrees ($\sim 8.8~ \rm Mpc^{2}$). The
FAUST completeness limit is $m_{\rm UV}\sim12.2$ (Cohen et al. \cite{cohen}), significantly lower than the FOCA magnitude 
limit: $m_{\rm UV}\sim18.5$.
However combining the two UV catalogs we hope to constrain the shape of the UV luminosity function across 7 magnitudes.
We use the FAUST observations for $m_{\rm UV}<12.2$ and the FOCA observations for  $m_{\rm UV}\geq12.2$.
To account for the different response function of FAUST and FOCA filters we
transform the UV magnitudes taken by FAUST at $\rm1650\AA$ 
assuming a constant color index: UV(2000) = UV(1650) + 0.2 mag 
(Deharveng et al. \cite{deharveng94}, \cite{deharveng}). We think however that this difference
does not bias the galaxy populations selected by the two experiments.
The estimated error on the UV magnitudes is 0.3 mag in general, but it ranges from 0.2 mag for bright galaxies, 
to 0.5 mag for faint sources observed in frames with larger than average calibration uncertainties.
The UV emission associated with bright galaxies is generally clumpy, thus it has been obtained integrating
the flux over the galaxy optical extension, determined at the surface brightness of 25 $\rm mag~arcsec^{-2}$ in the B-band. 
The spatial resolution of the UV observations is 20 arcsec and 4 arcmin for FOCA and FAUST respectively.
The astrometric accuracy is therefore insufficient for unambiguously discriminating between stars and galaxies.
To overcome this limitation, we cross-correlate the UV catalogs with the deepest optical catalogs of galaxies available: 
the Virgo Cluster Catalog (VCC, Binggeli et al. \cite{binggeli}), complete to $m_{\rm B}\sim18$, for the Virgo cluster and the $r'$ band catalog by 
Iglesias-Paramo et al. (\cite{jorge03}), complete to $m_{\rm r'}\sim20$, for Coma and Abell 1367. 
We used as matching radius the spatial resolution of each observation. 
In case of multiple identifications we select the galaxy closest to the UV position.
The resultant UV selected sample is composed of 156 galaxies in Virgo, 140 galaxies in Coma and 133 galaxies in Abell 1367.

\section{The UV luminosity functions}
\begin{figure}[t]
\centering
\includegraphics[width=8.5cm]{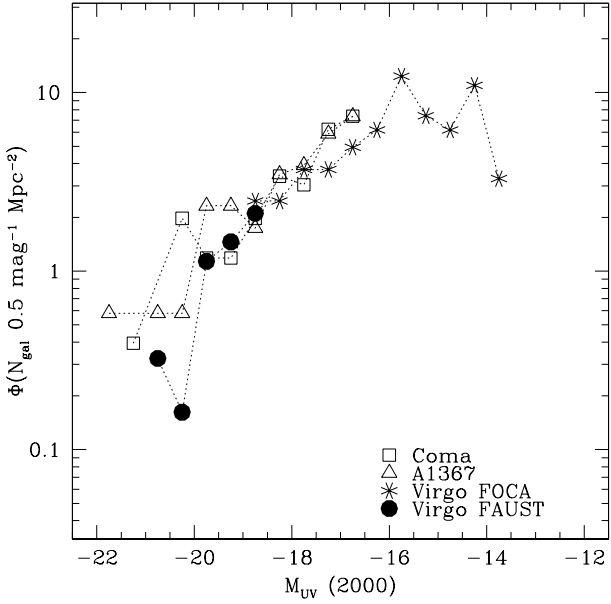}
\caption{The UV luminosity functions for the four analyzed data sets.}
\label{lfuv}
\end{figure}

Contrary to the VCC catalog, the Coma and A1367 $r'$ catalogs used for star/galaxy discrimination do not cover all the area observed 
by FOCA but only the cluster cores.
This reduces our analysis to an area of $\sim$ 1~ square degrees ($\sim 2.6~ \rm Mpc^{2}$) in Coma and $\sim$ 0.7 ~ square degrees 
($\sim 1.8~ \rm Mpc^{2}$) in Abell 1367.\\
Including new spectroscopic observations (Cortese et al., in preparation), the redshift completeness of the UV selected sample reaches the 
65\%  in Abell 1367, the 79\% in Coma and the 83\%  in Virgo. 
The redshift completeness per bin of magnitude of each cluster is listed in Table 1.
\begin{table}
\caption{Integral redshift completeness in bin of 0.5 magnitudes.}
\label{tab1} \[ 
\begin{array}{cccc}
\hline 
\noalign{\smallskip}   
  &\rm Redshift &\rm completeness & \\
M_{\rm UV}\leq & \rm Virgo & \rm Coma & \rm Abell1367  \\
\noalign{\smallskip} 
\hline 
\noalign{\smallskip}
-21.75 &  -    &  -	& 100\% \\
-21.25 &  -    & 100\%  & 100\%	\\
-20.75 & 100\% & 100\%	& 100\% \\
-20.25 & 100\% & 100\%  & 100\% \\
-19.75 &  92\% & 100\%  & 100\% \\
-19.25 &  95\% & 100\%  & 100\% \\
-18.75 &  97\% & 100\%  & 100\% \\
-18.25 &  97\% & 97\%	& 100\% \\
-17.75 &  97\% & 95\%	& 95\%	\\
-17.25 &  98\% & 84\%	& 80\%	\\
-16.75 &  98\% & 79\%	& 65\%	\\
\noalign{\smallskip} 
\hline 
\end{array} \]
\end{table}
\begin{table}[b]
    \caption[]{The completeness-corrected differential number of galaxies per bin of magnitude}
      \label{counts}
     $$ 
        \begin{tabular}{lcccc}
        \hline
        \noalign{\smallskip}
 $M_{\rm UV}$ & \multicolumn{4}{c}{$N_{i}$} \\
 mag & Virgo & Virgo & Coma & Abell~1367 \\
   & (Faust) & (Foca) & & \\
        \noalign{\smallskip}
        \hline
        \noalign{\smallskip}
 $-$21.75  &  0  &  0  &     0  &      1 \\
 $-$21.25  &  0  &  0  &     1  &      0 \\
 $-$20.75  &  2  &  0  &     0  &      1 \\
 $-$20.25  &  1  &  0  &     5  &      1 \\
 $-$19.75  &  7  &  0  &     3  &      4 \\
 $-$19.25  &  9  &  0  &     3  &      4 \\
 $-$18.75  & 13  &  2  &     5  &      3 \\
 $-$18.25  &  0  &  2  &  8.6  &      6 \\
 $-$17.75  &  0  &  3  &  7.7  &   6.7 \\
 $-$17.25  &  0  &  3  & 15.8  &  10.1 \\
 $-$16.75  &  0  &  4  & 18.6  &  12.7 \\	
	\noalign{\smallskip}
        \hline
        \end{tabular}
     $$ 
\end{table}
We remark that for $M_{\rm UV} \leq -16.5$ (corresponding to the FOCA magnitude limit in Coma and Abell1367), 
the redshift completeness of the Virgo cluster sample is 98\%.\\
As discussed by Cortese et al. (\cite{cortese}), the general UV galaxy counts (Milliard et al. \cite{milliard92}) are 
uncertain and cannot be used to obtain a reliable subtraction of the background contribution from the cluster counts.
Therefore, in order to compute the cluster LF, we use the statistical approach recently proposed by De Propris et al.(\cite{depropris}) 
and Mobasher et al.(\cite{mobasher}). 
We assume that the UV spectroscopic sample is 'representative', in the sense that the fraction of galaxies that are 
cluster members is the same in the (incomplete) spectroscopic sample as in the (complete) photometric 
sample.
For each magnitude bin $i$ we count the number of cluster members $N_M$, the number of galaxies with a measured recessional 
velocity $N_Z$ and the total number of galaxies $N_T$.
The completeness-corrected number of cluster members in each bin is:
\begin{equation}
\label{lf}
N_i = \frac{N_M N_T}{N_Z}
\end{equation}
$N_T$ is a Poisson variable, and $N_M$ is a binomial variable (the number of successes in $N_Z$ trials with probability $N_M$/$N_Z$).
Therefore the errors associated with $N_i$ are given by:
\begin{equation}
\label{err.sing}
\frac{\delta^{2}N_i}{N_i^{2}} = \frac{1}{N_T} + \frac{1}{N_M} - \frac{1}{N_Z}
\end{equation}
The completeness-corrected number of cluster members obtained from (\ref{lf}) are given in Table 2 
and the luminosity 
functions for the four studied samples are shown in Fig.\ref{lfuv}.
The two different datasets used for the Virgo cluster have only one 
magnitude bin ($M_{\rm UV} = -18.75$) overlap. In this bin the two LFs are in agreement 
and there is no indication that a change of slope occurs.
We thus feel comfortable 
combining them into a composite Virgo UV luminosity function across 7 magnitudes.\\
In order to determine whether the LFs of the three clusters are in agreement we perform 
a two-sample $\chi^{2}$ test. 
We obtain $P(\chi^{2}\geq\chi^{2}_{\rm obs})\sim $82\%  for the Virgo and Abell1367 LFs, 
$P(\chi^{2}\geq\chi^{2}_{\rm obs})\sim $87\% for the Virgo and the Coma cluster LFs and 
$P(\chi^{2}\geq\chi^{2}_{\rm obs})\sim $98\% for the Coma and Abell1367 LFs, pointing out 
that the three LFs are in fair agreement within their completeness limits.

\begin{figure}[t]
\centering
\includegraphics[width=8.5cm]{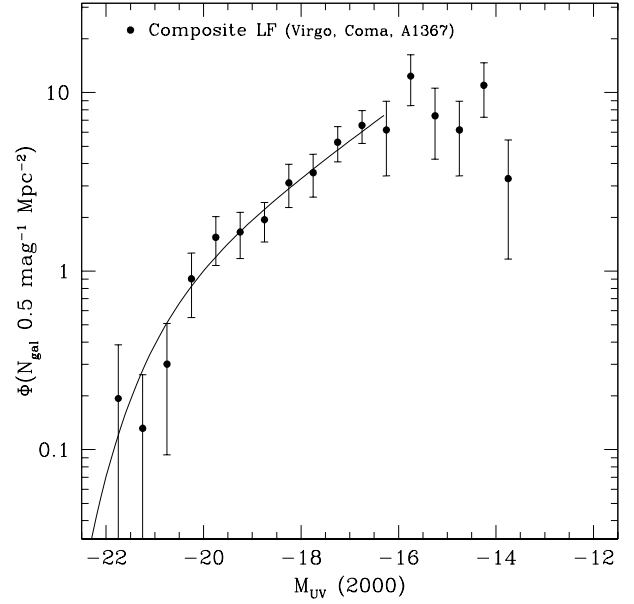}
\caption{The composite UV luminosity function of 3 nearby clusters. The solid line represents the best Schechter fit to the data for 
$M_{\rm UV} \leq -16.5$.}
\label{composite}
\end{figure}
\begin{figure}[t]
\centering
\includegraphics[width=8.5cm]{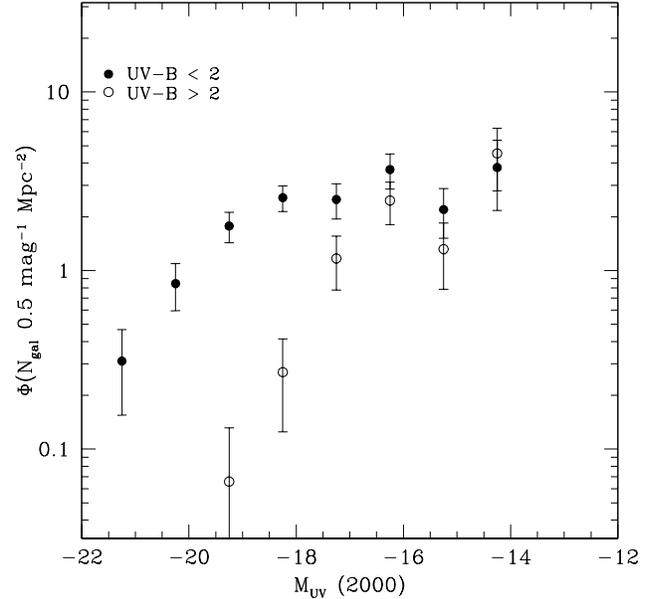}
\caption{The UV bi-variate composite luminosity functions of nearby clusters. Red ($\rm UV-B>2$) and blue ($\rm UV-B<2$) galaxies are 
indicated with empty and filled circles respectively.}
\label{bivariate}
\end{figure}

\subsection{The composite cluster luminosity function}
The uncertainties of each individual cluster luminosity function are too large to fit a complete Schechter (\cite{schechter})
function to the data and compare it with the field UV LF.
However combining the three data-sets analyzed in this paper we compute the UV composite luminosity function 
of 3 nearby clusters. 
The composite LF is obtained following Colless (\cite{colless}), by summing galaxies in absolute magnitude bins and 
scaling by the area covered in each cluster.
The number of galaxies in the $j$th absolute magnitude bin of the composite LF ($N_{cj}$) is given by:
\begin{equation}
\label{comp}
N_{cj} = \frac{1}{m_j} \sum_{i}\frac{N_{ij}}{A_i}
\end{equation}
where $N_{ij}$ is the completeness-corrected number of galaxies in the $j$th bin of the $i$th cluster, $A_i$ is 
the area surveyed in the $i$th cluster and $m_j$ is the number of clusters contributing to the $j$th bin.
The errors in $N_{ij}$  are computed according to:
\begin{equation}
\delta N_{cj} = \frac{1}{m_j} \Big[\sum_{i}\Big(\frac{\delta N_{ij}}{A_i}\Big)^{2}\Big]^{1/2}
\end{equation}
where $\delta N_{ij}$ is the error in the $j$th bin of the $i$th cluster determined in (\ref{err.sing}).
The weight associated to each cluster is computed according to the surveyed area, instead of the number of 
galaxies brighter than a given magnitude, as used by Colless (\cite{colless}).\\
The UV composite luminosity function is given in Fig.\ref{composite} in the full magnitude range. 
However since for magnitudes 
fainter than $ M_{\rm UV}\sim-16.5$ the only available data are the Virgo FOCA observations,
we fit the composite luminosity function with the Schechter functional 
form (Schechter \cite{schechter}):
\begin{displaymath} 
\phi(M_{\rm UV}) = 0.4~ \ln10~\phi^*~10^{0.4(M^*-M_{\rm UV})(\alpha+1)}~e^{-10^{0.4(M^*-M_{\rm UV})}}
\end{displaymath} 
only for $ M_{\rm UV}\leq-16.5$, that is the completeness limit in Coma and Abell 1367. 
The resulting Schechter parameters are $M_{\rm UV}^*$ = $-$20.75 $\pm$ 0.40 and $\alpha$ = $-$1.50 $\pm$ 0.10.
The faint-end slope is consistent within 1 $\sigma$ with the lower limit for Coma and A1367 recently proposed by Cortese et al.(\cite{cortese}),
but significantly flatter than the slope $\rm \alpha\sim-2.0,-2.2$ found for Coma by Andreon (\cite{andreon}), 
suggesting that this very steep luminosity function was due to an underestimate of the density of background galaxies.
\begin{figure}[t]
\centering
\includegraphics[width=8.5cm]{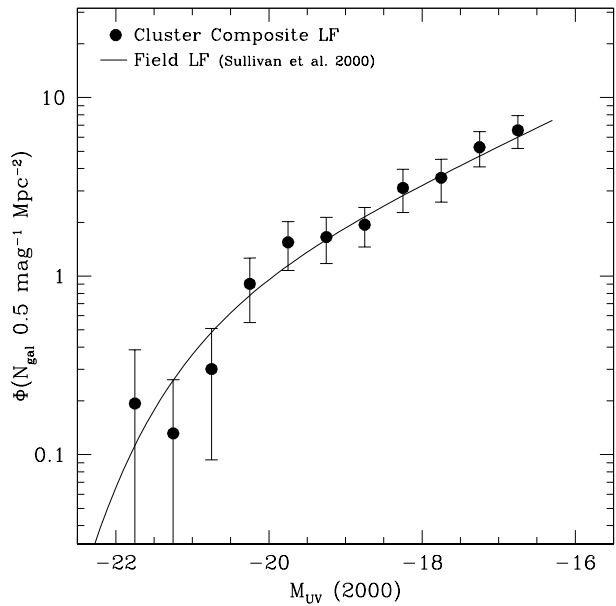}
\caption{The cluster and the field UV luminosity functions. The composite cluster LF is given with filled circles. 
The solid line indicates the best Schechter fit of the field LF of Sullivan et al.
(\cite{sullivan}). The normalization is such that the two LFs match at $M_{\rm UV} \sim -19.25$.}
\label{field}
\end{figure}

\section{Discussion}
Although the UV($\rm 2000~\AA$) radiation is dominated by young stars of intermediate masses 
(2$<$M$<$5$M_{\sun}$, Boselli et al. \cite{boselli}), it is frequently detected also in early-type galaxies with no recent 
star formation episodes (Deharveng et al. \cite{deharveng}). 
Unfortunately we have no morphological (or spectral) classification for all the UV selected galaxies in order to separate the 
contribution of late and early type galaxies. 
However, based on the spectral energy distributions computed by Gavazzi et al.(\cite{gav02}), 
we can use the total color $\rm UV-B$, available for the 94\% of galaxies in our sample, 
to discriminate between red elliptical ($\rm UV-B>2$) and blue spiral ($\rm UV-B<2$) galaxies.
B magnitudes are taken from the VCC (Binggeli et al. \cite{binggeli}), the Godwin et al. (\cite{godwin83}) catalog 
and the Godwin \& Peach (\cite{godwin82}) catalog for Virgo, Coma and Abell 1367 respectively.\\ 
The bi-variate composite luminosity function derived for galaxies of known $\rm UV-B$ color is shown in Fig.\ref{bivariate}. 
It shows that the star forming galaxies dominate the UV LF for $M_{\rm UV}\leq -18$, as Donas et al. (\cite{donas91}) 
concluded for the first time. 
Conversely, for $M_{\rm UV}\geq -17.5$, the number of red and blue galaxies 
is approximately the same, pointing out that, at low luminosities, the UV emission must be ascribed 
not only to star formation episodes but also to Post-Asymptotic Giant Branch 
(PAGB) low mass stars in early type galaxies (Deharveng et al. \cite{deharveng}).
Similarly, if we restrict the analysis to the fraction ($\sim 50$ \%) of objects with known morphological type,
we find that late-types (Sa or later) dominate at bright UV luminosities, while 
early-type objects contribute at the faint UV levels.
Since Virgo and Abell1367 are spiral-rich clusters while Coma is spiral-poor,
one might expect that the LFs of the three clusters obtained combining all types should have different shapes,
contrary to the observations. 
The point is that the combined LF of the two types is dominated, at high UV luminosity by the
spiral component, while at low luminosity early- and late-type galaxies contribute similarly. 
The UV LF of the spiral component are similar in the three clusters. 
At faint UV luminosities also the number density of early-type galaxies 
is approximately the same in the three clusters.
Only at relatively high UV luminosity the number density
of early-type galaxies in the Coma cluster exceeds significantly that of the other two clusters, but it is
still much lower than the one of the late-type component. Therefore the LF obtained by combining
early- with late-type galaxies results approximately the same in the three clusters.\\
The cluster composite luminosity function has identical slope and similar $M^*$ as the UV  
luminosity function computed by Sullivan et al.
(\cite{sullivan}) for the field: $M_{\rm UV}^*$ = $-$21.21 $\pm$ 0.13, $\alpha$ = $-$1.51 $\pm$ 0.10, 
as shown in Fig. \ref{field}.
This result points in the same direction as recent studies of cluster galaxies carried out in $\rm H\alpha$ 
(Iglesias$-$Paramo 
et al. \cite{jorge02}) and B-bands (De Propris et al. \cite{depropris}).
They find that the LFs of star forming galaxies in clusters and in the field have the same shape,
contrary to early type galaxies in clusters that have a brighter and steeper LF than their field counterparts 
(De Propris et al. \cite{depropris}). 
This indicates that, whatever 
mechanism (i.e. ram pressure, tidal interaction, galaxy harassment) quenches/enhances 
the star formation activity in late-type cluster galaxies, it influences similarly the
giant and the dwarf components, so that the shape of their LF results unchanged and only the normalization
is modified.

\begin{acknowledgements}      
This work could not have been completed without the use of the NASA/IPAC Extragalactic Database (NED) which 
is operated by the Jet Propulsion Laboratory, Caltech under contract with NASA.
We also made use of the GOLDmine Database, operated by the Universit\'{a} degli Studi di Milano-Bicocca. 
\end{acknowledgements}

\end{document}